\documentclass[aps,pre,floats, onecolumn,showpacs,superscriptaddress]{revtex4}

\usepackage{graphicx,epsfig}
\usepackage{graphics,dcolumn,bm,epic, eepic,fleqn,float}
\usepackage{amssymb,amsmath,multirow,rotate,xcolor}

\begin{document}

\preprint{APS/123-QED}

\title{Cooperation in costly-access environments}

\author{Hugo P\'erez-Mart\'inez}
\affiliation{Department of Condensed Matter Physics, University of Zaragoza, 50009, Zaragoza, Spain.}
\author{Carlos Gracia-L\'azaro}%
\affiliation{Institute for Biocomputation and Physics of Complex Systems, University of Zaragoza, 50018, Zaragoza, Spain.}%
\author{Fabio Dercole}%
\affiliation{Department of Electronics, Information, and Bioengineering, Politecnico di Milano, 20133, Milano, Italy.}%
\author{Yamir Moreno}%
\affiliation{Institute for Biocomputation and Physics of Complex Systems, University of Zaragoza, 50018, Zaragoza, Spain.}%
\affiliation{Department of Theoretical Physics. University of Zaragoza, 50005, Zaragoza, Spain.}%
\affiliation{ISI Foundation, Turin, Italy.}%

\date{\today}

\begin{abstract}
Understanding cooperative behavior in biological and social systems constitutes a scientific challenge, being the object of intense research over the past decades. Many mechanisms have been proposed to explain the presence and persistence of cooperation in those systems, showing that there is no unique explanation, as different scenarios have different possible driving forces. In this paper, we propose a model to study situations in which voluntary participation involves an access cost to the cooperative interaction, besides the cost associated with cooperation. The proposed Costly-Access Prisoner's Dilemma, a symmetric donation game with voluntary and costly participation, breaks the symmetry between abstainers and participants of the Voluntary Prisoner's Dilemma. A mean-field approach shows that, in well-mixed populations, the dynamic always leads the system to abstention.
However, depending on the return parameter, numerical simulations in structured populations display an alternating behavior between mono-strategic, multi-stable, and coexistence phases. This behavior is fully explained through a theoretical analysis of the strategic motifs, the transitions being determined by the change in stability of those motifs.
\end{abstract}

\maketitle


\section{\label{sec:intro}Introduction}

Cooperative behavior constitutes an open question in science being one of the current key scientific challenges \cite{pennisi2005did}. Although many works have dealt with it since the original contributions by Trivers \cite{trivers1971evolution}, Axelrod and Hamilton \cite{axelrod1981evolution}, there is not a universal explanation for the emergence and fixation of cooperation in scenarios where selfish behavior involves a higher individual benefit. Several mechanisms have been proposed to explain it \cite{nowak2006five}, including kin selection \cite{hamilton1964genetical,molina2019intergenerational}, direct \cite{trivers1971evolution} or indirect reciprocity \cite{roberts2008evolution,xia2020effect}, group selection \cite{traulsen2006evolution}, and network reciprocity \cite{nowak1992evolutionary,santos2005scale,gomez2007dynamical,gracia2012heterogeneous,dercole2019direct,dellarossa2020reciprocity}.

In this regard, the Evolutionary Game Theory \cite{robert1984evolution,sigmund2010calculus}
provides a fruitful theoretical framework for studying cooperative behavior \cite{VonNeumann1944, Myerson1997}, including  cooperation in structured populations \cite{szabo2007evolutionary,perc2010coevolutionary}. Within this framework, the Prisoner's Dilemma (PD) constitutes the most representative and widely studied game for modeling cooperative behavior evolution  \cite{Rapoport1965,szabo2005phase,eguiluz2005cooperation,szolnoki2007cooperation,perc2008social,vilone2014social,hilbe2015partners,javarone2016statistical}. The PD models pairwise situations where altruistic behavior involves a cost (such as time, energy, or risks) that produces a benefit for the opponent, while uncooperative behavior (defection) provides no benefit for the counterpart at no cost. When the PD is evolutionarily played in unstructured populations, the dynamic leads to a fully-defector population, cooperation needing incentive mechanisms to persist \cite{nowak2006five}.

In addition to the above-mentioned mechanisms, voluntary participation may be an effective way of promoting cooperation \cite{hauert2002volunteering}. Furthermore, it constitutes a setup to model situations in which agents may decide to participate or not in a potential interaction. In these setups, in addition to cooperation and defection, the loner strategy is introduced to account for risk aversion. The introduction of loners induces cyclic strategies' dominance, which cause circular orbits in well-mixed populations \cite{hauert2002replicator} and strategical domains in networks \cite{szabo2002phase}. In the case of pairwise interactions, both Voluntary Prisoner's Dilemma (VPD) \cite{szabo2002evolutionary} and Prisoner’s Dilemma with Exit Option  \cite{vanberg1992rationality,arend2005modeling,phelan2005using} model situations in which agents are allowed to decide between avoiding participation in a PD (loner strategy) and participating by choosing either to cooperate or defect. VPD displays cooperative behavior persistence in structured populations, from coexistence domains in lattices to different oscillation trajectories types in random networks \cite{szabo2002evolutionary,hauert2005vpd,chen2010evolutionary,guo2017environment}.

In the VPD, any pairwise interaction involving at least a loner provides the same payoff for both players. Considering the game in terms of costs of participation to the underlying PD, cooperators and defectors have to pay a cost, or a fee, only when they face each other and the PD interaction does take place; otherwise, no one gets anything. This interpretation is formally obtained by subtracting the loaner payoff to all entries of the payoff matrix in the original VPD \cite{szabo2002evolutionary}.
It models situations involving a fixed cost from the PD interaction (interaction cost in the following), not including the cost of the altruistic act of cooperation, and no alternative source of payoff for loaners.

In this work, we propose a modification of the optional PD to model situations in which the willingness to participate in the cooperative setup involves a cost, even if the opponent player refuses the interaction.
The motivation is to study those scenarios where the access to the interaction place (whether physical or virtual) entails an expense, such as transport costs \cite{hampton1999environmental}, entry fees \cite{buchanan1965economic,imerman1993application},
or time investment \cite{rogelberg2012wasted},
which can be avoided by refusing interaction. Although participation costs in PD have been studied \cite{masuda2007participation}, there is a lack of theoretical studies on the effect of entry fees on non-mandatory participation. The proposed model corresponds to a symmetric donation game with voluntary and costly participation. By imposing a participation fee, we break the symmetry of the VPD through a payoffs difference between the player that refuses to interact (hereafter,  the abstainer or loner) and her counterpart (the attendant, i.e., cooperator or defector). The proposed two-person game, hereafter Costly-Access Prisoner's Dilemma (CAPD), has three strategies: cooperation, defection, and abstention. While abstention does not involve any payoff (neither benefit nor fee), defection and cooperation entail a participation fee besides the cooperation cost associated with the latter. Note that, in the proposed CAPD, the players willing to participate in the underlying PD pay the participation fee, regardless of whether PD takes place or not. Conversely, in the VPD, the participation fee is paid only if the PD takes place. We can interpret the CAPD as a risky version of the VPD, as showing up to participate involves a risk: this risk breaks a symmetry of the VPD.

First, through a mean-field approach, we show that abstention is the only attractor of the system in well-mixed populations. Next, we perform numerical simulations on different networks of multiple sizes, from Random Regular to Erd\"os-R\'enyi and Barab\'asi-Albert graphs. The numerical results show a rich scenario of transitions for increasing values of the
benefit-to-cost ratio of the underlying donation game (the enhancement factor of the cost of cooperation that represents the economic return of the game)
between mono-strategic, multi-stable, and coexistence phases.
Then, we perform a microscopic analysis that satisfactorily characterize those transitions by the thresholds that the changes in the stability of a limited set of motifs generate.

\section{\label{sec:theModel}The Model}

In this section, we define the proposed CAPD game and the dynamics. Let be a population of $n$ agents --the players-- endowed with a network structure. The nodes of the network represent the agents and the links the connections among them. The interaction between any pair of agents takes place through a PD with voluntary and costly participation. Specifically, agents are allowed to adopt one of the three available strategies: cooperation ($C$), defection ($D$), and abstention ($A$). Each agent takes one of the above strategies when playing with all her neighbors. If the agent decides to abstain, she does not pay nor receive anything. Otherwise, she must pay a participation fee $t$. Furthermore, cooperation has an additional cost $c$ (the contribution) to the participation fee, while defection does not entail additional cost. The counterpart of a cooperator receives $rc$, i.e., her partner's contribution $c$ multiplied by the enhancement factor $r$.

We will take $c=1$ without loss of generality. Once two players decide to participate (none is an abstainer), this formulation of the PD is equivalent to a Reciprocal Donation Game: In that game, each player is allowed to donate 
$c$; if she does, her counterpart receives $rc$.

In each interaction, the strategy of both agents determines their payoffs. 
Let $\sigma_i$ be the strategy of player $i$; $\sigma_i^T=(1,0,0),(0,1,0),(0,0,1)$ for an abstainer, cooperator and defector, respectively. The payoff obtained by player $i$ facing $j$ is given by $\sigma_i^T M \sigma_j$, where $M$ is the payoffs matrix:
\begin{align}
M=
\begin{pmatrix}
\;\;\;0 \;\;\;&\;\;\; 0 \;\;\;&\;\;\; 0\\
\;\;\;-t \;\;\;&\;\;\; r-t-1 \;\;\;&\;\;\; -t-1\\
\;\;\;-t \;\;\;&\;\;\; r-t \;\;\;&\;\;\; -t
\end{pmatrix}\nonumber
\end{align}

A cooperator or a defector who faces an abstainer must pay the participation fee $t$ because she has presented himself to play, but since the game is not played (the abstainer does not appear), cooperator will not pay the cooperation cost $c=1$. Furthermore, the highest payoff corresponds to a defector facing (and therefore exploiting) a cooperator, while the lowest to a cooperator facing a defector. 

Note that, by introducing the participation fee, we broke a symmetry of VPD in which both agents obtain the same payoff when one of them abstains. Furthermore, removing strategy $A$ the game reduces to the classical PD with the Reciprocal Donation Game formulation.

\subsection*{Dynamics}
Initially, strategies are equiprobably distributed among the agents unless otherwise specified.
At each time step:
\begin{enumerate}
    \item Each agent plays with the same strategy ($A$, $C$, or $D$) with all her neighbors. The payoff $\Pi_i$ of an agent, $i$, will be the sum of those obtained when playing with all her neighbors.
    \item Once all the agents have played, they decide whether to keep their strategy in the next round or to change it. Specifically, each agent $i$ randomly selects a neighbor $j$ and compares their payoffs. Agent $i$ adopts $j$'s strategy with a probability given by:
    \begin{equation}
P_{ij}=\frac{1}{1+\exp(\frac{\Pi_i-\Pi_j}{T})} \;\;.
\label{eq:fermiUpdatingRule}
\end{equation}

\end{enumerate}

According to equation \ref{eq:fermiUpdatingRule}, when the $j$'s payoff is significantly higher than $i$'s one, agent $i$ will almost certainly imitate $j$'s strategy. Conversely, if it is significantly lower, agent $i$ will keep hers. However, if both payoffs are similar, agent $i$ may choose the strategy with worse results (in any case, with a probability lower than 1/2)\cite{szabo1998evolutionary}. This mechanism represents the possibility that agents make irrational decisions, or make mistakes when estimating payoffs. The parameter $T$ stands for the temperature or noise. The higher the value of $T$, the higher the noise and, correspondingly, the probability of choosing the lower payoff strategy. In evolutionary terms, i.e., regarding the selection intensity, high $T$ corresponds to weak selection pressure. 

\section{\label{sec:MF}Mean-field approach}

Before studying the behavior of the system in networks, we perform a mean-field approximation. This approximation will provide information about the behavior of the system in a well-mixed population. In this approach, we assume that we can identify the strategies in the neighborhood of any agent with the composition of the complete system. Then, given the Markovian character of the updating rule, the dynamics may be described through the total fractions of agents of each strategy present in the system at a given time $\rho_A, \rho_C, \rho_D$, where $\rho_A+\rho_C+\rho_D=1$. 

The total payoff $\Pi_x$ of an $x$-strategist of degree $k$ is given by:
\begin{equation}
\Pi_x = k (\rho_x M_{xx} +  \rho_y M_{xy} + \rho_z M_{xz})\;\;,
\label{eq:totalPayoffMeanField}
\end{equation}
where $x$ stands for any strategy and the other two indices, $y$ and $z$, for the other two strategies.

Eqs. \ref{eq:fermiUpdatingRule} and \ref{eq:totalPayoffMeanField} provide the imitation probabilities. Consider a network of fixed degree $k$, and let us assume that focal agent of strategy $x$ has chosen a neighbor of strategy $y$ ($x,y=A,C,D;\;x\neq y$). The probability that the focal agent will adopt strategy $y$ is given by: 
\begin{equation}
    P_{xy}=\frac{1}{1+exp\left [kT^{-1} \sum\limits_{i\in \{A,C,D\}} \rho_i (M_{xi}-M_{yi})\right ]}\;\;.
    \label{Eq:TransitionProbabilities}
\end{equation}

The temporal evolution of the frequencies for each strategy are:

\begin{align}
\dot{\rho}_A =& \rho_A \left [ \rho_C (P_{CA} - P_{AC}) + \rho_D (P_{DA} - P_{AD} ) \right ]\;, \nonumber \\
\dot{\rho}_C =& \rho_C \left [ \rho_D (P_{DC} - P_{CD}) + \rho_A (P_{AC} - P_{CA} ) \right ] \;, \nonumber \\ 
\dot{\rho}_D =& \rho_C \left [ \rho_D (P_{CD} - P_{DC}) + \rho_A (P_{AD} - P_{DA} ) \right ]\;. \nonumber\\ 
\label{eq:flows}
\end{align}

Note that the sum of the flows is zero: $\dot{\rho}_A+\dot{\rho}_C-\dot{\rho}_D=0$. 
The system will be in equilibrium if $\dot{\rho}_A=\dot{\rho}_C=\dot{\rho}_D=0$. Given the normalization condition $\rho_A+\rho_C+\rho_D=1$, the system evolution can be characterized through the frequency of two strategies and their corresponding flows.

\begin{figure*}
\centering
 \includegraphics[width=0.9\columnwidth]{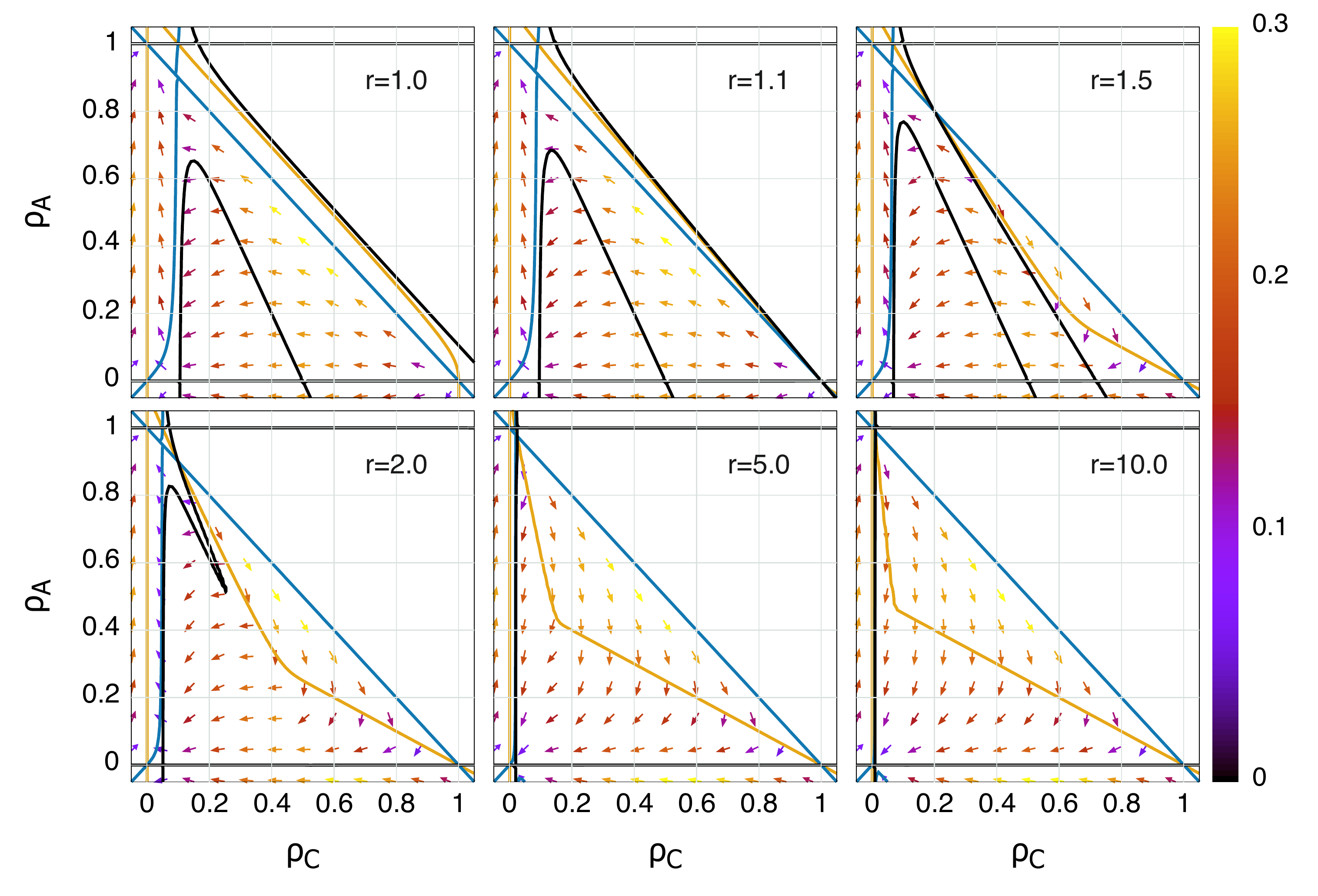}
 \caption{\textbf{Phase portraits.} 
 Flow diagrams for different values of $r$. Blue, yellow and black lines correspond to $\dot{\rho}_D=0$, $\dot{\rho}_C=0$, $\dot{\rho}_A=0$  nullclines, respectively. The intersections of nullclines show the fixed points. The arrows indicate the direction of the flows, and their color, their speed. Other parameters values are $T=0.1, k = 4, t = 0.1$. See the main text for further details.}
\label{fig:phasePortraits}
\end{figure*}

Figure \ref{fig:phasePortraits} displays the flow diagrams for different values of $r$, arrows represent the trajectories and lines the nullclines: blue, yellow, and black lines correspond to $\dot{\rho}_D=0$, $\dot{\rho}_C=0$ and $\dot{\rho}_A=0$, respectively. The intersection of nullclines determines a fixed point (note that the intersection of two nullclines implies the intersection of the third one, i.e., a T-junction). The analysis of these T-junctions displays the existence of several fixed points. In the first place, the three vertices of the simplex, corresponding to the Full-$C$, Full-$D$, and Full-$A$ states, are the trivial fixed points of the system, all of them absorbing states. A third fixed point also appears on the hypotenuse, corresponding to the absence of $D$ agents, which is unstable.

The three border regions ${\rho}_C=0$, ${\rho}_D=0$, and ${\rho}_A=0$
correspond to invariant manifolds. Out of these regions, i.e., in the interior of the simplex, the flows are such that the system always ends up in the Full-$A$ state for any value of $r$. Note that this Full-$A$ state constitutes the only Nash equilibrium of the model. However, we can observe that there is a behavior change when $r>c+t=1+t$. The center-top panel corresponds to the limit $r=c+t$ ($r=1.1$ for the plot's parameters): for this value of the parameter $r$, the nullcline $\dot{\rho}_C=0$ becomes tangent to the other two nullclines at the Full-$C$ point, and for higher values of $r$ ($r>c+t$), it falls into the phase space, giving rise to a flow region that points in the direction of decreasing $\rho_A$. This fact will be analyzed in subsequent sections beyond this mean-field approach: Although in infinite systems all the inner trajectories eventually leave this area and end at the Full-$A$ point, in finite systems, a trajectory with stochastic effects could lead to the extinction of $A$-strategists, confining the dynamics to a PD system in which the only stable point is Full-$D$.

Finally, the only fixed point involving coexistent strategies, which always complies with $\rho_D=0$, is unstable. 
To prove this, first note that, in a fixed point, the payoffs of all non-zero frequency strategies must be equal so that the inflows and outflows of each strategists' set equalize. The asymmetry in the payoff matrix between strategies $C$ and $D$, defectors obtaining higher payoff than cooperators for the same fraction of cooperating neighbors, avoids fixed points with the coexistence of cooperators an defectors. As in the absence of cooperators, abstainers obtain higher payoff than defectors, fixed points involving two coexistent strategies are restricted to the $\rho_D=0$ invariant manifold. Within this manifold, a fixed point in the $0<\rho_A<1$ region has to satisfy $P_{AC}=P_{CA}$.  
Regarding stability, in the immediate vicinity of that point, an increment of $\rho_C$ (resp., $\rho_A$) involves  $P_{AC}>P_{CA}$ (resp., $P_{CA}>P_{AC}$); from Equation \ref{eq:flows} follows that $\dot{\rho}_C>0$ (resp., $\dot{\rho}_A>0$), implying a perturbation growing, which proves that the fixed point is unstable.

\section{\label{sec:NumericalResults}Numerical results on networks}

\subsection*{\label{sec:RRN}Random regular network}

In the previously discussed mean-field approximation,
we assumed that, in any given moment, the fractions of strategists in the neighborhood of a node are the corresponding fractions of the system, i.e., we neglected the local fluctuations. In order to contextualize the numerical results with those obtained with the mean-field approach, we perform the spatial simulations in Random Regular Networks (RRN). In an RRN, all the nodes have the same degree, such that each node is randomly connected to other $k$ nodes. 

\begin{figure}[t]
 \includegraphics[width=\columnwidth]{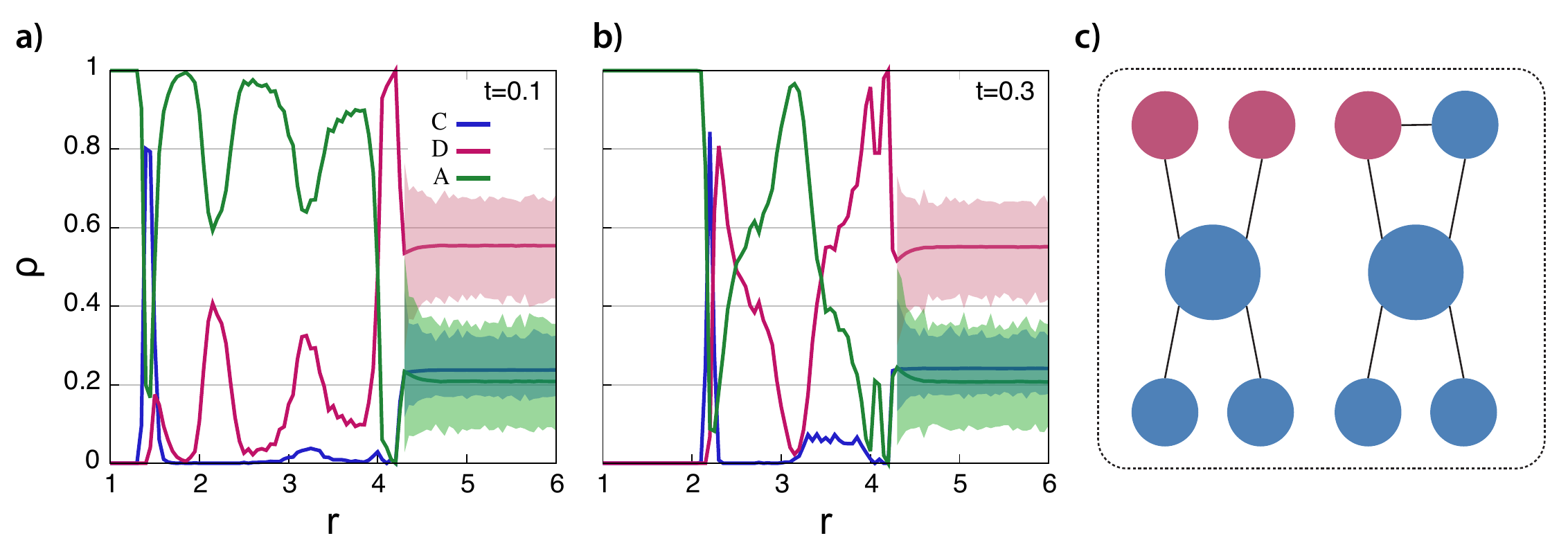}
 \caption{\textbf{Numerical results in a RRN for T=0.1.} Fraction of cooperators (C), defectors (D), and abstainers (A) versus the enhancement factor $r$ for a random regular network in presence of noise ($T>0$). For $r<r_c \sim 4.2$, all the simulations end on one-strategy absorbing states, and therefore the results correspond to the fraction of simulations that end up in each state. Conversely, for $r>r_c$, all the simulations end on three-coexistent-strategies states. Shadowed regions indicate the maximum oscillation sizes of each strategy in the coexistence case. Panel \textbf{A)} corresponds to a participation fee of $t=0.1$; Panel \textbf{B)} to $t=0.3$. In both panels, the network size is $N=10^4$; $k=4$, $T=0.1$. Each point in the one-strategy dominance region corresponds to the average over 1000 independent simulations, while the points in the coexistence region correspond to the average over 100 simulations. \textbf{c)} Motifs affecting threshold in $r=r_c$. See the text for further details.}
\label{fig:RRN}
\end{figure}

Figure \ref{fig:RRN} displays the numerical results for an RRN of $N=10^4$ nodes and degree $k=4$; $T=0.1$. Panel A corresponds to a participation fee of $t=0.1$ and Panel B to $t=0.3$. Both panels show the final fractions of each strategy as a function of the enhancement factor $r$, for an equiprobable initial distribution of strategists ($\rho_C(t=0)=\rho_D(t=0)=1/3$). The system shows three regions characterized by different behaviors and bounded by two thresholds. For very low values of the enhancement factor, all the simulations end in the full-$A$ state, as predicted by the mean-field approach. Nevertheless, there is a value $r_{\min}=1.4$ for the enhancement factor $r$ beyond which not all executions end in the full-$A$ state. The second transition happens at a critical value $r_c\sim 4.2$. For $r<r_c$, all simulations end in a one-strategy absorbing state, the fractions representing the number of simulations that end in each one of them. Conversely, for $r>r_c$, all the simulations end in a three-strategies coexistence state. Furthermore, the fractions of strategists in this latter state are independent of $r$. To summarize, the system exhibits three phases: a first full-$A$ mono-strategic phase for $r<r_{\min}$; a second multi-stable phase consisting of three possible final states (full-$C$, full-$D$, and full-$A$); and a third coexistence phase characterized by an inner three-strategies stationary state. The main discrepancy with the mean-field approach is that, for $r>r_{\min}$, some simulations may end in full-$C$ or full-$D$ states, or even in three-strategies coexistence states. Therefore, these effects must be because of the underlying network structure. 

To delve into the effect of the structure on the dynamics, let us first analyze the system stability. If into a full-$C$ population, a $D$-agent is introduced, the latter will obtain a higher payoff by exploiting the rest; therefore, strategy $D$ will spread, and the state will end up being full-$D$. On the other hand, a full-$D$ system, which supports negative payoffs, will be unstable against an $A$-agent. Finally, a full-$A$ population will be unstable against a pair of $C$-strategists for a sufficiently high value of r, as shown in the following paragraph. As a corollary, for the system to end up in a one-strategy state, its dominated strategy has to end its predatory one: the domination by a strategy needs the prior extermination of that one that invades it. For example, for the dynamics to lead to a full-$C$ state, $A$-agents must have dominated all the $D$-agents first.

We characterize now the thresholds $r_{\min}$ and $r_c$. First, $r_{\min}$ corresponds to the value for which a cooperator connected to another cooperator obtains the same payoff as an abstainer. Then, in the early (randomly distributed) stages, cooperators will avoid invasion. Note that payoffs can be expressed as:

\begin{align}
\Pi_A =& 0, \nonumber \\
\Pi_C =& N_C^C r-(N_C^C + N_D^C)-kt\;, \nonumber \\
\Pi_D =& N_C^D r-kt\;, \nonumber \\
\label{eq:payoffsStructure}
\end{align}
where $N_X^Y$ is the number of $X$-strategists in the neighborhood of an $Y$-strategist. Taking $k=4$, the above-mentioned condition $\Pi_C(N_C^C=1)=\Pi_A$ provides $r_{\min} = 4t+1$, which fits the limit of the always-full-$A$ region in both panels A-B of Fig. \ref{fig:RRN}, namely,
$r_{\min} (t=0.1)= 1.4$ (Panel A) and $r_{\min}(t=0.3)= 2.2$ (Panel B). Before this threshold, there were no configurations aside from $C$-triangles (i.e. three $C$-agents that neighbor each other, very improbable in RRN) and some other more complex structures that could obtain payments higher than 0, and thus, survival of $C$ and $D$ strategies was almost impossible. The cooperation peak that appears when $r\simeq r_{\min}$ will be discussed later.

Second, $r_c$ marks the onset of coexistence between the three strategies. Note that a three-strategies equilibrium configuration requires one or more cooperators clusters arranged so that defectors located at the cluster border cannot invade them. Beyond those, there has to be a set of $A$-agents, to which those defectors who cannot find a cooperator to exploit will imitate. As the three-strategies coexistence equilibrium requires the stability of the $C$-clusters, a condition for the stable coexistence is that $C$-agents at the cluster's border must obtain a higher payoff than their neighboring $D$-agents. Motifs in Fig. \ref{fig:RRN}C show the minimum configurations around a $C$-agent (blue circles) for that stability, (Note that, in those cases, the neighboring D-agents (magenta circles) cannot have any adjacent $C$-agents apart from those explicitly shown). In those motifs, the limit for the central $C$-agent not being invaded by D-strategy, $\Pi_C=\Pi_D$, provides the value $r_c=4$. The transition occurs for a slightly higher value of r as an effect of temperature $T>0$: for $r\gtrsim 4$, there is a non-zero probability for a border's $C$-agent to imitate a $D$-agent with a lower payoff, unbalancing the system and destroying the coexistence. Note that, regardless of the value of $r$, a $C$-agent will be unstable against a $D$-agent with the same number of neighboring $C$. Therefore, an increase in the value of $r$ beyond the transition will not affect the equilibrium, as there are no additional motifs that can become stable and change the survival capabilities of $C$-clusters. This fact explains why the equilibrium composition of the system does not change for $r\gg r_c$.

\begin{figure*}[t]
\centering
 \includegraphics[width=.6667\textwidth]{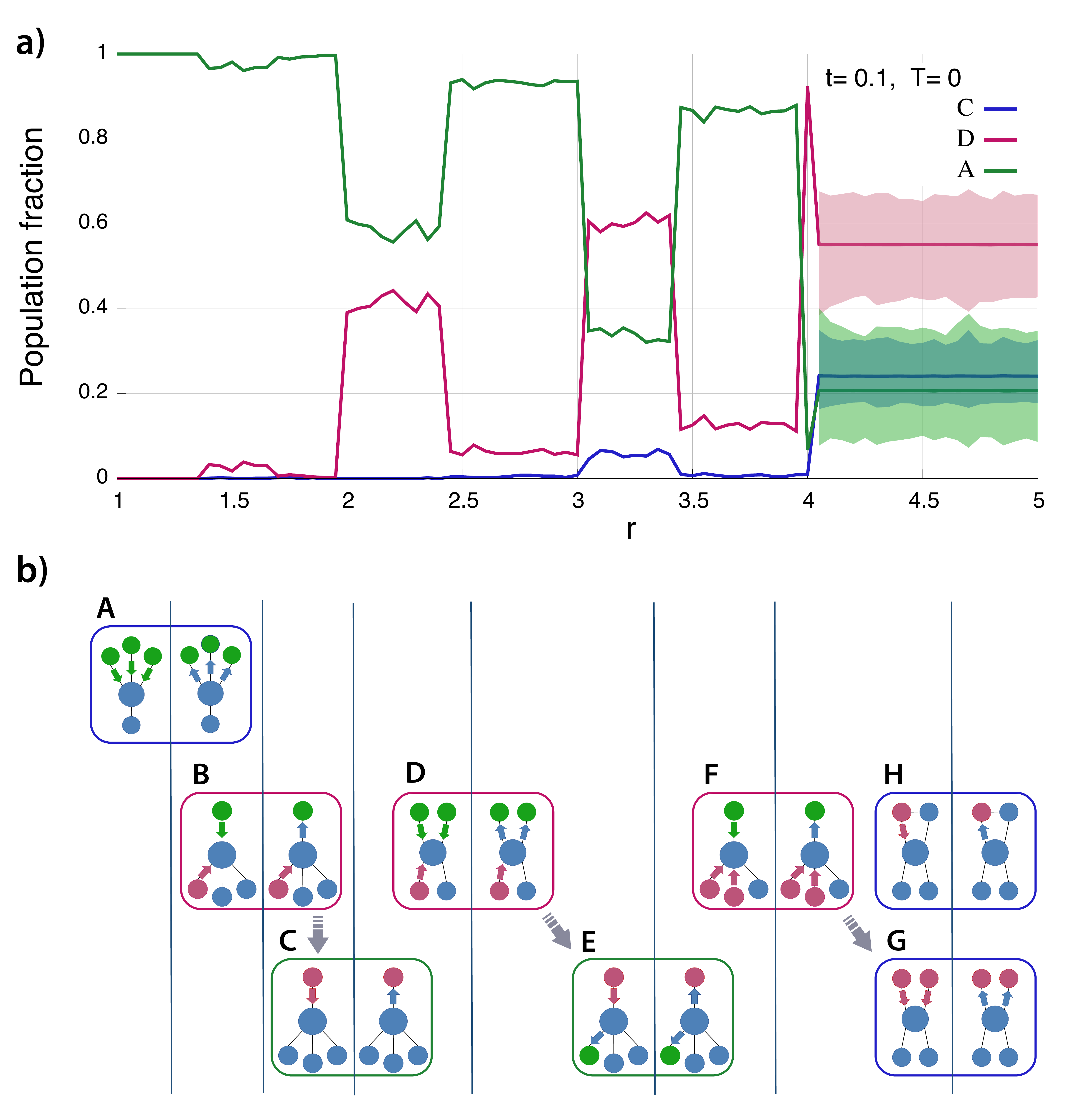}
 \caption{\textbf{Numerical results for a RRN, T=0.} \textbf{a) }Fraction of cooperators (C), defectors (D), and abstainers (A) versus the enhancement factor $r$ in a random regular network for an updating rule according to a step function ($T=0$). Shadowed regions indicate the maximum oscillation sizes of each strategy in the three-strategies coexistence case. The network size is $n=10^4$ with fixed degree $k=4$, and participation fee $t=0.1$. Each point in the one-strategy dominance region corresponds to the average over 1000 independent simulations, while the points in the coexistence region correspond to the average over 100 simulations. \textbf{b)} Diagrams a to h correspond to the configuration transitions. Each diagram indicates a dominance transition for the corresponding motifs regarding the propagation or extinction of the central node. In these diagrams, neighbors of cooperators' $D$-neighbors are not cooperators. The purple arrows indicate the system evolution when the central $C$-node invades an $A$-neighbor.}
\label{fig:RRN.T0}
\end{figure*}

To study the behavior inside the region $r_{\min}<r<r_c$ it is convenient to remove the stochastic effects introduced by the Fermi update rule, which dilutes the location of the transitions making the results difficult to interpret. To that end, we numerically study the dynamics for T=0, which involves a step function in Eq. \ref{eq:fermiUpdatingRule}, i.e., agents only imitate neighbors with higher payoffs than theirs. Note that, as the choice of the neighbor to imitate remains random, the dynamics preserve some degree of stochasticity.

Panel A of Fig. \ref{fig:RRN.T0} shows the results for $t=0.1$ and $T=0$. Panel B displays a series of motifs concerning the stability of a cooperator (blue circle) surrounded by i) other cooperators, ii) defectors (magenta), or iii) abstainers (green). Within these diagrams, the left (resp. right) motif represents the dominance for lower (higher) values of $r$ than the corresponding threshold (grey, vertical lines). The links' arrows indicate strategy spreading: If blue, cooperation can propagate in that direction; in contrast, a green (magenta) arrow pointing towards the central node means that abstainer (defector) neighbor can invade it. It should be noted that for $T=0$, quantitative payoffs differences do not affect the dynamics but rather which one among two agents has a higher payoff. That gives rise to plateaus in the final fractions as an $r$ increment is only relevant if it implies a change in the dominance hierarchy. Conversely, for $T>0$, the payoff differences have quantitative relevance, with a non-zero probability of imitating a neighbor with a lower (but close) payoff, which engenders smoother transitions and peak shapes. 

Starting with the first plateau shown in Fig. \ref{fig:RRN.T0}, for $r_{\min} = 1.4<r<1.7$ there is a significant probability that the system ends up in a full-$D$ state. For $r>r_{\min}$, cooperators can spread as shown in Diagram a, and then, there is a chance that they will exterminate abstainers, leading the system to a full-$D$ state. However, according to the results shown in Figure \ref{fig:RRN.T0}, the increase in the number of simulations that end up in full-$D$ states is marginal, which means that the extermination of $C$ is still quite more probable than the extermination of $A$. The reason behind this can be found in the fact that there are still very few $C$ structures that can withstand a $D$ invasion or propagate while neighboring a $D$-agent, so any $C$-cluster connected to a $D$-agent will eventually disappear. The main effect of the quick propagation of $C$ is then that all independent $C$-clusters will become connected to another $C$-cluster contaminated with neighboring $D$-agents before they disappear, and will thus be exterminated before completely invading $A$.

Note that by using the Fermi rule with $T = 0.1$, we had before that $P_{CA} = P_{AC}$ when $r\simeq r_{\min}$ in the configuration of two cooperators surrounded by abstainers. Thus, the propagation of these $C$-clusters was mostly slow and stochastic, and all $C$-clusters with neighboring $D$-agents would decay before they could link up with other $C$-clusters not harassed by $D$, eventually reaching a full-$C$ state as observed in panels A and B of Figure \ref{fig:RRN}. For slightly higher $r$ values, the propagation of cooperators became faster, reaching structures with defectors before the latter ones disappeared and recovering the full-$A$ behavior previously explained.

This behavior changes for $r>1.7$, where a new path for the propagation of $C$-clusters contaminated by $D$-agents emerges. In this new scenario, the $C$-nodes of the configuration shown in Diagram b become stable against abstainers, involving a faster spreading of cooperation. Again, the rapid expansion of cooperators that coexist at a disadvantage with defectors allows them to reach other cooperation clusters previously undisturbed. Then, they will fall victim to defection, leading eventually to their extermination and ending almost always in full-$A$ states. This phenomenon did not happen for $r<1.7$ as the clusters of coexistence between cooperators and defectors were unable to propagate, making the extinction of $A$-agents by uncontaminated $C$-clusters more probable and slightly increasing the number of simulations that ended up in full-$D$ states. Note that the ability of $C$-clusters to propagate over $A$ is not the only key feature, but also the ability of $D$-agents to propagate over neighboring $C$, as both effects determine the probability of extermination of $C$ or $A$-agents.
Counterintuitively, a hampered propagation of $D$ over $C$ gives rise to bigger population fractions of $C$, which in turn facilitates their quick invasion by $D$, that is, benefiting one strategy entails the promotion of the strategy that predates it \cite{hauert2005vpd}. Note that the larger the total fraction of $C$, the more probable it is for a $D$ agent to neighbor two or more $C$-agents, which means that its propagation over $C$ can reactivate. This means that all thresholds usually bring two competing effects to the table: they increase the propagation speed of cooperation (more probable extinction of $A$, and an increase on the number of Full-$D$ states), but as a result, they provide a solid foundation for the quick propagation of $D$ when $C$ is a majority, allowing for the extinction of $C$ to happen before the extinction of $A$. As we will see later, network size also plays a crucial role because it increases the survival probabilities of all strategies.

The relation between propagation abilities of $C$ and $D$ comes to effect when considering the value of $r$ in which the invasion of $C$-clusters by $D$ is hampered by some $C$-motifs. This is actually the case in the next threshold, $r=2$, when the central $C$-agent of the configuration shown in Diagram C matches her payment with that of the $D$-neighbor, $\Pi_D = \Pi_A$, and is then able to resist invasion. This means that the extermination of $C$ is way less probable than before, as the propagation of $D$ over $C$ is effectively closed in a number of cases. Note that this configuration is achieved, for example, when the central $C$-agent of Diagram B propagates over her $A$-neighbor, which offsets the detrimental effect that the quick propagation of contaminated $C$-clusters had before as now they can quickly become decontaminated. As a result, extermination of $A$ becomes common and the fraction of simulations that end up in full-$D$ states increase drastically. However, extermination of $C$ by $D$ is still possible, as configurations like that shown in Diagram C are only temporarily stable against $D$. For example, neighboring $C$-agents of the central $C$ can be invaded by their own neighbors before the $D$-agent disappears, causing the collapse of the whole structure. If $A$ can survive long enough so that $C$ reaches a sufficiently high population fraction that allows for the invasion by $D$ to happen, then we could see a decrease in the final $D$-fraction rather than an increase.

The same pattern of extinctions arises in order to explain the rest of the thresholds up to $r=r_c$. For all values of $r$ in which a new way of propagation of $C$-clusters that cannot defend against neighboring $D$-agents appear (Diagrams B, D and F), there is a sudden drop in the fraction of simulations that end up in full-$D$ states, as the extermination of $C$-agents is promoted. In turn, for values of $r$ in which some ways of propagation of $D$-agents are blocked (Diagrams C and E, which can be reached by propagating $C$ from Diagrams B and D), the survival of $C$-agents increases and so does the probability of $A$-agents disappearing, causing a rise in the number of simulations that end up in full-$D$ states.

\begin{figure*}
\centering
 \includegraphics[width=.6667\textwidth]{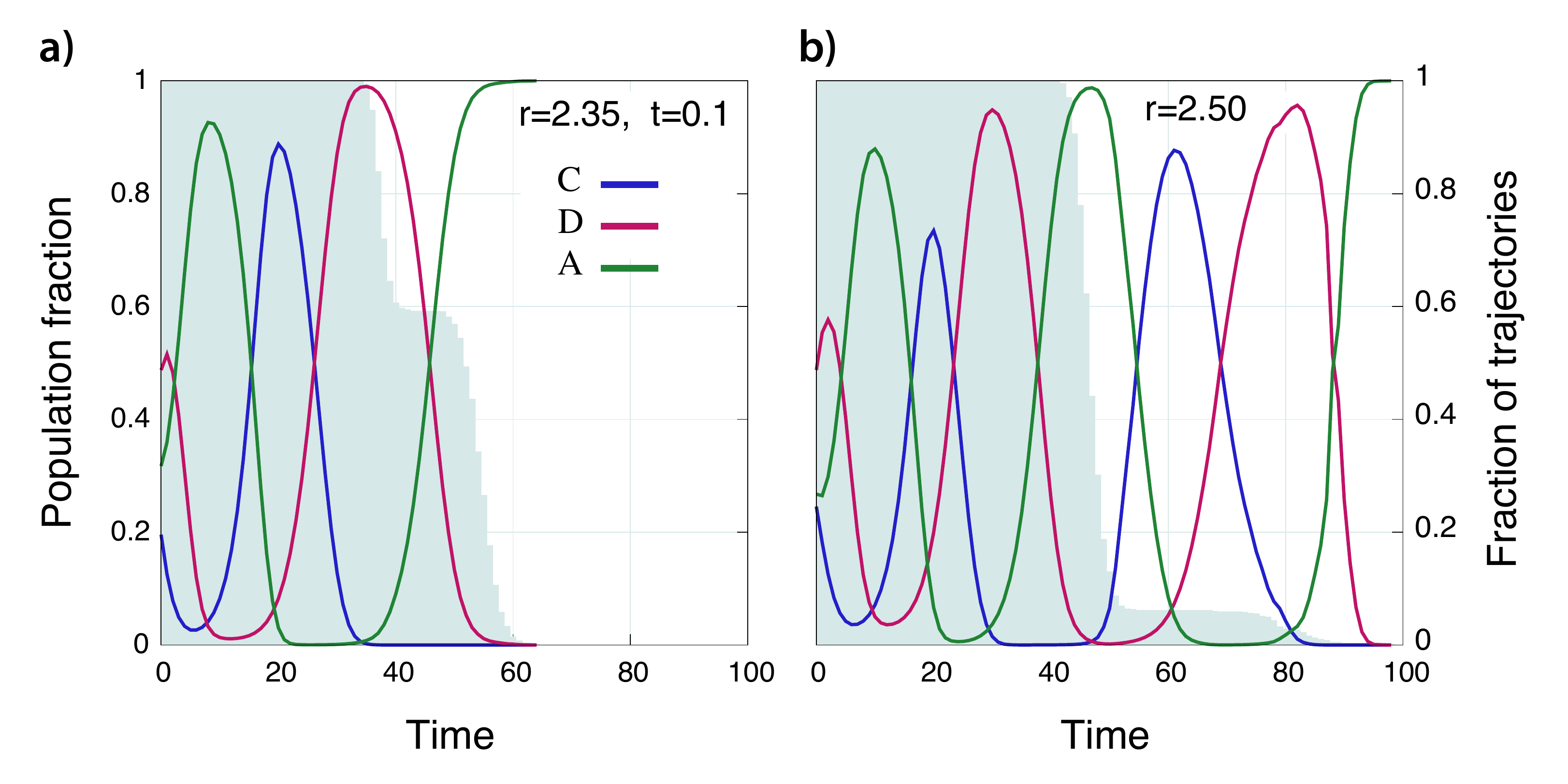}
 \caption{\textbf{Time evolution for a RRN.} Frequencies of the total fractions of each strategy as a function of time in a regular random network of $N=10^4$ nodes and degree $k=4$. Panel a) shows the results for $r=2.35$, and b) for $r=2.5$. Solid lines correspond to the average over 1000 simulations, and grey shadow to the fraction of trajectories remaining active at a certain time. In all the plots, $T=0$ and $t=0.1$. See the text for further details.} \label{fig:evolution.RRN.T0}
\end{figure*}

To illustrate the effect of the motifs' dominance transitions on the dynamics and the resulting final state, Fig. \ref{fig:evolution.RRN.T0}  displays the time evolution of the strategies frequency for two values of the enhancement factor: $r=2.35$ (panel A), and $r=2.5$ (panel B). In these plots, it is shown that population fractions oscillate, dominant strategy changing cyclically between abstention, cooperation, and defection, while the oscillation frequency is remarkably homogeneous. We can also see the time periods in which some trajectories come to an end, understood as the survival of only one strategy. Those trajectories reaching an absorbing state at a given time are removed from the plot at that point, the gray shadow representing the fraction of active simulations: a decrease on the fraction of trajectories at a certain time period means that those trajectories have ended in states fully dominated by the strategy whose fraction is the greatest at that same period. Therefore, the number of realizations that end in states dominated by a given strategy (gray shadow) matches the fractions shown in Figure \ref{fig:RRN.T0}. Panel A of Fig. \ref{fig:evolution.RRN.T0} corresponds to the region $2\leq r < 2.4$ in Fig. \ref{fig:RRN.T0}A, with significant fractions of both full-$A$ and full-$D$ final states. On the other hand, panel C ($r=2.5$) corresponds to a defection's channel opening, specifically the point $r=2.4$ beyond which a non-isolated cooperator connected to two abstainers and susceptible of being invaded by a defector is allowed to invade abstainers (Diagram D in Fig. \ref{fig:RRN.T0}B). From this value, $r\geq 2.4$, as shown in panel C, the new stable motif hinders the propagation of $C$, which is not able to exterminate $A$ around $t=20$, and is in turn exterminated around $t=30$, leading finally to a massive increase in the total fraction of full-$A$ simulations.

\subsection*{\label{sec:NetworkSize}Effect of network size}

As we said before, strategy extinctions are a finite size effect, as larger networks have a higher probability of stable motifs appearing. Moreover, increasing network size could result in local desynchronization of the trajectories presented before or the occasional appearance of certain temporal higher-order motifs that could withstand extermination by the majority strategy. Both effects would result in increased survival capabilities of all three strategies. Considering that the final proportions of each strategy result from the race between the extinction of $C$ and $A$, network size could play a determinant role in the final population fractions that we observe in the absorbing state region.

\begin{figure*}
\centering
 \includegraphics[width=\textwidth]{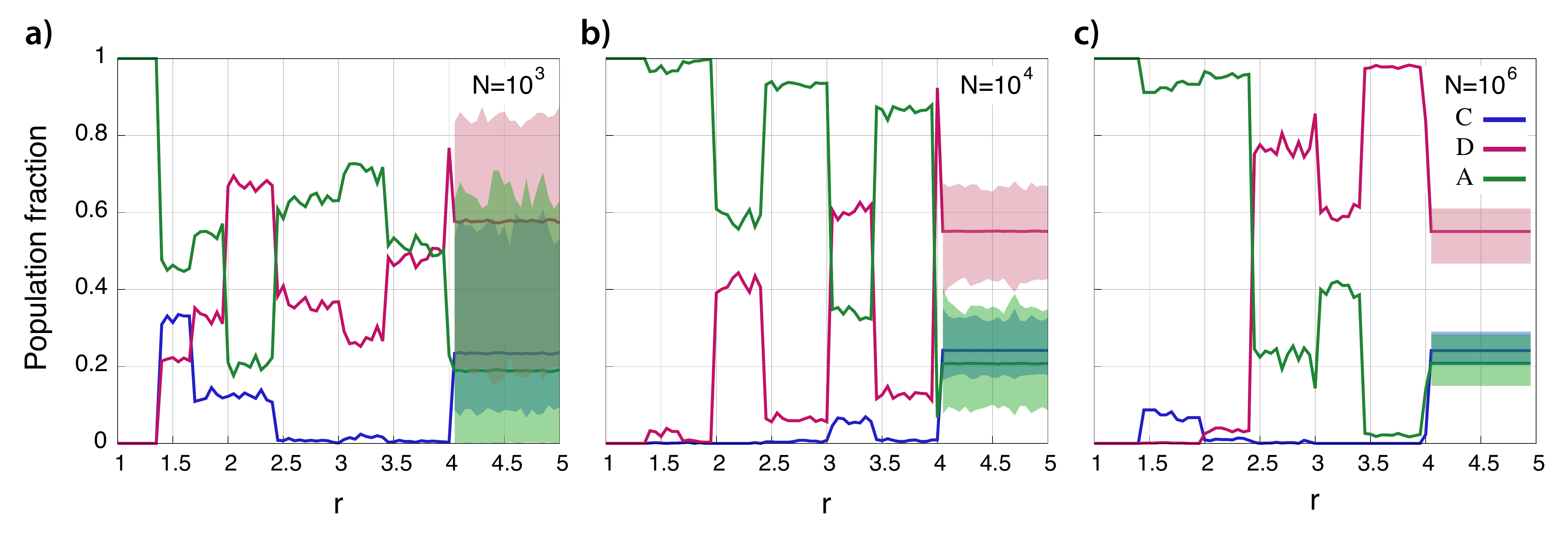}
 \caption{\textbf{Numerical results for RRN of different sizes, T=0, $\mathbf{t=0.1}$.} Fraction of cooperators (C), defectors (D), and abstainers (A) versus the enhancement factor $r$ in random regular networks for increasing network sizes. Shadowed regions indicate the maximum oscillation sizes of each strategy in the three-strategies coexistence case. Each point in the one-strategy dominance region corresponds to the average over 1000 independent simulations, while the points in the coexistence region correspond to the average over 100 simulations. \textbf{a)} Network of $N=10^3$ nodes. \textbf{b)} Network of $N=10^4$ nodes. Note that this panel corresponds to Panel a) of Figure \ref{fig:RRN.T0}. \textbf{c)} Network of $N=10^6$ nodes.} 
 \label{fig:pop_fracs_N}
\end{figure*}

To test this, we perform additional simulations for various network sizes. In Figure \ref{fig:pop_fracs_N} we present the results for $N=10^3$, $N=10^4$ and $N=10^6$. It is shown that network size has an important effect on the final proportion of simulations that end up dominated by each strategy, thus proving its influence on the survival probabilities. It is also clear that thresholds related to motifs are still present and give rise to abrupt transitions in final proportions. Explanations presented before remain valid, but the dominant effect, that is, the first strategy subject of extermination, depends on network size.

\begin{figure*}
\centering
 \includegraphics[width=.8\textwidth]{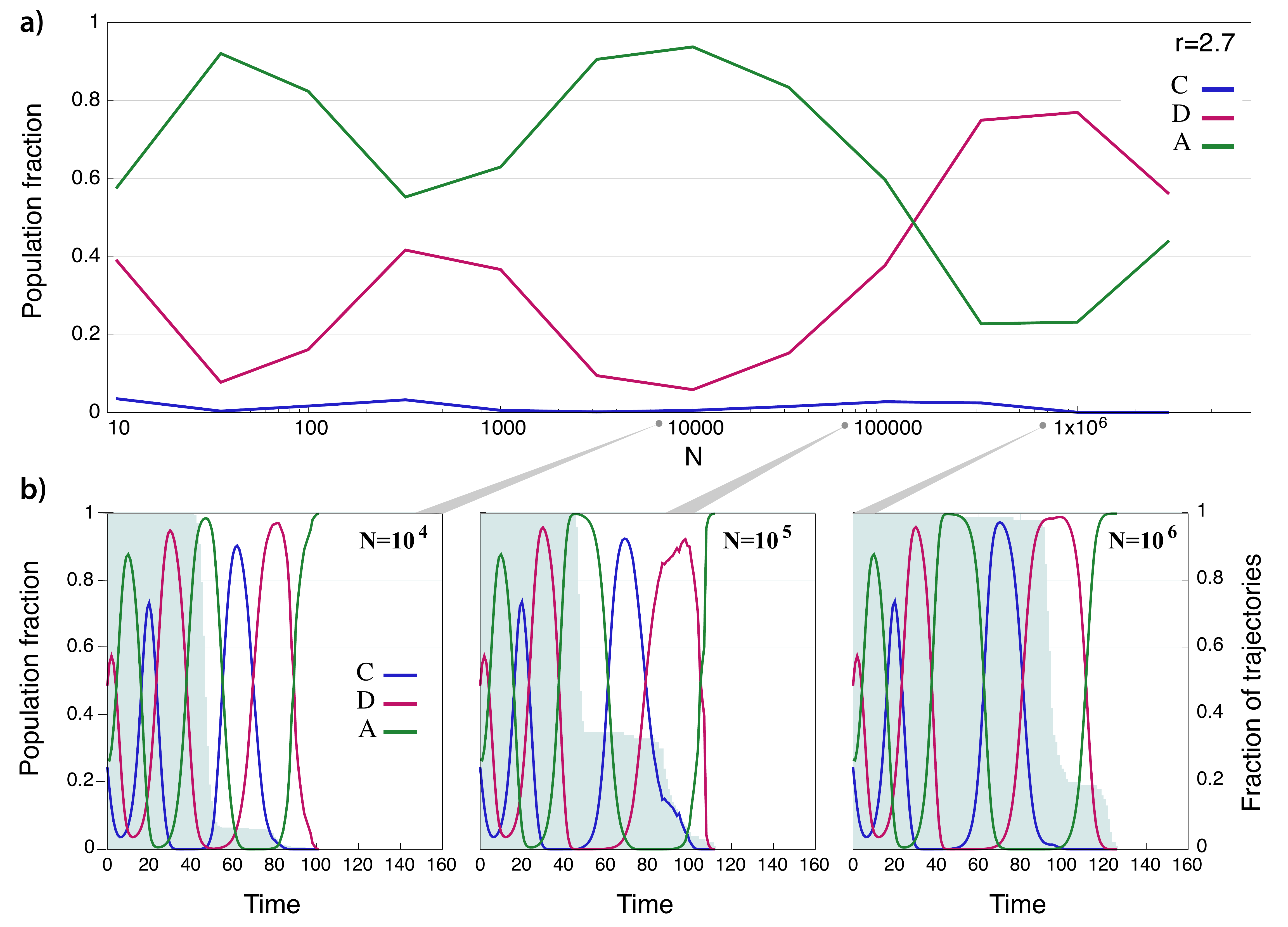}
 \caption{\textbf{Numerical results for fixed $r$ and increasing size.} \textbf{a)} Fraction of cooperators (C), defectors (D), and abstainers (A) versus the Network size $N$ in random regular networks for $r=2.7$. Each point corresponds to the average over 1000 independent simulations. \textbf{b)} Averaged trajectories for $r=2.7$ and three different network sizes: $N=10^4$, $N=10^5$ and $N=10^6$. Each trajectory correspond to the average of 1000 independent simulations, and the grey shadow to the fraction of trajectories remaining active at a certain time.} 
  \label{fig:pop_fracs_trajs}
\end{figure*}

To understand this dependence, we represent the final population fractions as a function of network size for $r=2.70$ in Figure \ref{fig:pop_fracs_trajs} where oscillations can clearly be seen. Their origin lies in the oscillating behavior of trajectories, which remains for all network sizes. To prove this, we represent in panel B averaged trajectories for three different network sizes. It can be seen that the oscillation amplitude and frequency are almost unaffected. Moreover, the trajectories minima reach very low values, which means that oscillations between different strategies must occur at the same time and that there must not be any local desynchronization in each individual trajectory. This rules out desynchronization as a cause for changing survival capabilities. However, it is also clear that bigger networks result in longer trajectories, which means that there is indeed an effect on the survival chance. In this respect, most trajectories for $N=10^4$ finish around t=$50$ because the extinction of $C$ is common around $t=40$, but as the network grows, this extinction becomes less probable, and other extinction mechanisms become relevant. For $N=10^5$, a fraction of trajectories end around $t=90$ because of the extinction of $A$ around $t=75$, and for $N=10^6$, this mechanism becomes dominant as almost all trajectories survive past the first extinction threshold. Now, some simulations are able to survive the extinction of $A$ in $t=75$, and thus, the extinction of $C$ becomes relevant again, this time around $t=100$. This is the clearest example of the competing effects on the survival of each strategy that motifs' thresholds have, as overcoming the ordeal imposed by the first of those effects brings the second one into account.

\subsection*{\label{sec:DegreeTopology}Effect of degree and network topology}

Results and discussions in previous section 
were related to a constant degree $k=4$. Although the influence of the stability structures on the dynamics may be generalized to other situations, it is clear that those structures will be different; therefore, both connectivity and network structure will influence system behavior.

\begin{figure*}
\centering
 \includegraphics[width=\textwidth]{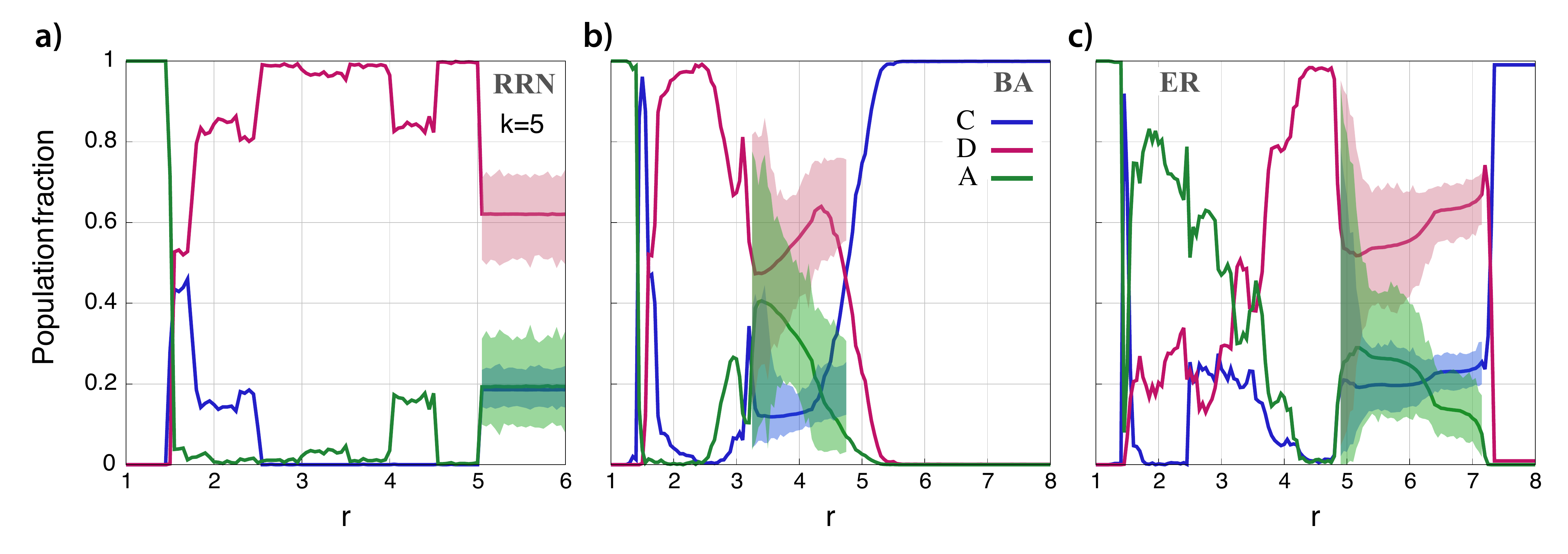}
 \caption{\textbf{Numerical results on different networks.} Fraction of cooperators (C), defectors (D), and abstainers (A) versus the enhancement factor $r$ for different networks. Panel \textbf{a)} corresponds to a random regular network of degree $k=5$; Panel \textbf{b)} to a Barab\'asi-Albert graph with mean degree $\langle k \rangle = 4$; and Panel \textbf{c)} to an Erd\"os-R\'enyi graph with $\langle k \rangle = 4$. In all the panels, $N=10^4$ nodes, $t=0.1$, and $T=0$.  Each point corresponds to the average over 1000 independent simulations. See the text for further details.
 } \label{fig:OtherNetworks}
\end{figure*}

Panel A of figure \ref{fig:OtherNetworks} shows the numerical results in an RRN of $N=10^4$ nodes and degree $k=5$, for a temperature $T=0$, a participation fee $t=0.1$, and a random equiprobable initial distribution of strategists. As the condition for the stable coexistence of the 3 strategies when $T=0$ is that $C$-agents at the cluster's border must obtain a higher payoff than their neighboring $D$-agents, this condition may be written as $r_c=k$. As discussed above for $k=4$, for $r<r_c$ all the simulations ended on one-strategy absorbing states, the numerical results corresponding to the average over the different simulations. By contrast, for $r>r_c$, i) all the simulations ended on 3-coexistent-strategies states and ii) strategists' frequencies are independent of $r$. The grounds leading to this result allow generalizing it for any fixed degree $k$. Nevertheless, the number and position of thresholds corresponding to the multi-stable phase $r_{\min}>r>r_c$ will depend on the degree. The number of motifs, and consequently the number of those interior thresholds and their multiplicity will increase with $k$.

Note the motif's analysis stability performed so far lies in the fact that the condition for a defector to obtain a higher payoff than an abstainer is to have, at least, a cooperator in its neighborhood. That condition preserves the no-dependence with $r$ of the defectors' stability against abstainers and is satisfied as long as $tk<r$. For larger values of $tk$, having one $C$-neighbor would no longer guarantee a positive payoff for the $D$-agent, new thresholds would appear, and the system's behavior would become more complex. For heterogeneous networks, the previous condition becomes $tk_{max}<r$, where $k_{max}$ is the maximum degree of the graph.

Let us now study the system behavior in complex networks with a heterogeneous degree distribution. Barab\'asi-Albert (BA) networks are random scale-free graphs generated through a preferential-attachment mechanism \cite{albert2002statistical}. These networks present power-law degree distributions, where few nodes (the hubs) have a very high degree. Panel B of Fig. \ref{fig:OtherNetworks} displays the numerical results of the model for a BA network of $N=10^4$ nodes, a mean degree of $\langle k\rangle = 4$, $T=0$, and $t=0.1$. The simulations show four phases: i) For $r<r_{\min} \sim 1.3$, the dynamic leads to a full-$A$ state; ii) for $r_{\min}<r<r_c\sim 3.2$, there is a multi-stable phase in which all the realizations end either in one-strategy states or in states almost dominated by one strategy; iii) for $r_c <r<r_{all-C}\sim 5$, the realizations end in three-strategies coexistent states, all the realizations for a given value of $r$ showing similar fractions of strategists; iv) finally, for $r>r_{all-C}\sim 5$, all realizations end in almost $C$-pure states. This new phase, characterized by the prevalence of cooperation, is a consequence of the hubs in the system. These hubs promote cooperation for high enough $r$ because once a hub adopts strategy $C$, a few cooperative neighbors are enough to avoid invasion and spread cooperation throughout the network. On the other hand, if a hub adopts strategy $D$, it will invade $C$-neighbors, decreasing its payoff, and, finally, it will become unstable against the remaining $C$-neighbors. This kind of behavior is well known and has been previously studied \cite{gomezgardenes2007dynamical,gomezgardenes2008natural,poncela2007robustness}.

However, the existence of high-degree nodes is not enough for this new almost-all-$C$ phase, as confirmed in tests in high-degree RRNs, where it is not present. As a necessary topological condition for the new phase, hub(s) must be surrounded by low-degree agents. Specifically, the condition for a $C$-hub to avoid invasion is to have more $C$-neighbors than the degree of any of its $D$-neighbors. Thus, while $D$-agents imitate a $C$-hub, they will unlikely achieve payoffs large enough to invade the $C$-hub, even when the latter has a modest number of $C$-neighbors.

Finally, we also studied the system in Erd\"os–R\'enyi networks (ER). ER networks are graphs in which nodes are equiprobably connected \cite{erdos2011evolution}. These graphs exhibit a binomial degree distribution, which tends to be a Poisson for large networks. Panel C of Fig. \ref{fig:OtherNetworks} shows the results for a ER network of $n=10000$ nodes, mean degree $\langle k\rangle = 4$, $T=0$, and participation fee $t=0.1$. As shown, ER networks exhibit a behavior qualitatively similar to BA ones, displaying the same four phases. Quantitatively, some thresholds between the four phases are shifted to higher values of $r$: $r_{\min} \sim 1.3$, $r_c \sim 5$, $r_{all-C} \sim 7.4$. These results confirm the fact that the arguments presented for BA networks apply qualitatively to any heterogeneous degree network.

\section{Conclusions}

In this paper, we have proposed a modification of the Prisoner's Dilemma, which we named Costly-Access PD (CAPD), by imposing a cost or fee to access the potential PD interaction and allowing agents to abstain from participating to avoid expenses. The introduced pairwise game has three strategies, namely cooperate, defect and abstain. It differs from the Voluntary Prisoner's Dilemma in its interpretation as a costly PD interaction. Our access fee breaks the symmetry of the VPD interaction fee, which is paid only if both players of the pairwise interaction eventually play the underlying PD. Our fee entails an \textit{a priori} cost to be covered by both cooperators and defectors to access the potential interaction, even if the interaction will not take place because refused by the other player. The idea behind this fee is to model situations where showing up to participate involves a risk, which can be avoided by refusing to interact.

In the absence of a network structure, the only evolutionary stable strategy is abstention, which coincides with the only Nash equilibrium of the system. We have studied the system through a mean-field approach, which resulted in the absence of inner fixed points, revealing the lack of coexistence stationary states. Furthermore, mean-field trajectories are such that any initial condition ends in a full-abstention state.

Extensive numerical simulations on graphs have shown that the network structure has a determining influence on the system dynamics. We have identified a series of strategic motifs whose stability thresholds determine the system behavior. The motifs' analysis allowed us to explain the different phases of the system and locate the critical values of the game return parameter $r$ that demarcate their boundaries. Also, it explains the jumps in the average strategies frequencies and provides the values of $r$ at which they take place. In particular, the system exhibits i) a first mono-strategic phase dominated by abstention, followed by ii) a multi-stable mono-strategic phase in which dynamics leads to one of the three absorbing states, iii) a three-strategies coexistence phase, and iv) for heterogeneous networks, a phase dominated by cooperation with a residual presence of defectors.

To conclude, by breaking the symmetry of the Voluntary Prisoners' Dilemma, we have presented a model that displays a rich phenomenology, with alternating stability switches driven by a single parameter. We believe this behavior will deserve the attention of physicists and mathematicians to be extrapolated to other systems.

\section*{Acknowledgments}

C. G. L and Y. M acknowledge partial support from the Government of Arag\'on and FEDER funds, Spain through grant ER36-20R to FENOL. Y. M also acknowledge partial support by MCIN/AEI and FEDER funds (grant PID2020-115800GB-I00). The funders had no role in study design, data collection and analysis, decision to publish, or preparation of the manuscript.

\bibliographystyle{ieeetr}

\bibliography{ParticipationFeeVPD}

\begin{thebibliography}{10}

\bibitem{pennisi2005did}
E.~Pennisi, ``How did cooperative behavior evolve?,'' {\em Science}, vol.~309,
  no.~5731, pp.~93--93, 2005.

\bibitem{trivers1971evolution}
R.~L. Trivers, ``The evolution of reciprocal altruism,'' {\em The Quarterly
  Review of Biology}, vol.~46, no.~1, pp.~35--57, 1971.

\bibitem{axelrod1981evolution}
R.~Axelrod and W.~D. Hamilton, ``The evolution of cooperation,'' {\em Science},
  vol.~211, no.~4489, pp.~1390--1396, 1981.

\bibitem{nowak2006five}
M.~A. Nowak, ``Five rules for the evolution of cooperation,'' {\em Science},
  vol.~314, no.~5805, pp.~1560--1563, 2006.

\bibitem{hamilton1964genetical}
W.~D. Hamilton, ``The genetical evolution of social behaviour. ii,'' {\em
  Journal of Theoretical Biology}, vol.~7, no.~1, pp.~17--52, 1964.

\bibitem{molina2019intergenerational}
J.~A. Molina, A.~Ferrer, J.~I. Gim{\'e}nez-Nadal, C.~Gracia-L{\'a}zaro,
  Y.~Moreno, and A.~Sanchez, ``Intergenerational cooperation within the
  household: a public good game with three generations,'' {\em Review of
  Economics of the Household}, vol.~17, no.~2, pp.~535--552, 2019.

\bibitem{roberts2008evolution}
G.~Roberts, ``Evolution of direct and indirect reciprocity,'' {\em Proceedings
  of the Royal Society B: Biological Sciences}, vol.~275, no.~1631,
  pp.~173--179, 2008.

\bibitem{xia2020effect}
C.~Xia, C.~Gracia-L{\'a}zaro, and Y.~Moreno, ``Effect of memory, intolerance,
  and second-order reputation on cooperation,'' {\em Chaos: An
  Interdisciplinary Journal of Nonlinear Science}, vol.~30, no.~6, p.~063122,
  2020.

\bibitem{traulsen2006evolution}
A.~Traulsen and M.~A. Nowak, ``Evolution of cooperation by multilevel
  selection,'' {\em Proceedings of the National Academy of Sciences}, vol.~103,
  no.~29, pp.~10952--10955, 2006.

\bibitem{nowak1992evolutionary}
M.~A. Nowak and R.~M. May, ``Evolutionary games and spatial chaos,'' {\em
  Nature}, vol.~359, no.~6398, pp.~826--829, 1992.

\bibitem{santos2005scale}
F.~C. Santos and J.~M. Pacheco, ``Scale-free networks provide a unifying
  framework for the emergence of cooperation,'' {\em Physical Review Letters},
  vol.~95, no.~9, p.~098104, 2005.

\bibitem{gomez2007dynamical}
J.~G{\'o}mez-Gardenes, M.~Campillo, L.~M. Flor{\'\i}a, and Y.~Moreno,
  ``Dynamical organization of cooperation in complex topologies,'' {\em
  Physical Review Letters}, vol.~98, no.~10, p.~108103, 2007.

\bibitem{gracia2012heterogeneous}
C.~Gracia-L{\'a}zaro, A.~Ferrer, G.~Ruiz, A.~Taranc{\'o}n, J.~A. Cuesta,
  A.~S{\'a}nchez, and Y.~Moreno, ``Heterogeneous networks do not promote
  cooperation when humans play a prisoner’s dilemma,'' {\em Proceedings of
  the National Academy of Sciences}, vol.~109, no.~32, pp.~12922--12926, 2012.

\bibitem{dercole2019direct}
F.~Dercole, F.~Della~Rossa, and C.~Piccardi, ``Direct reciprocity and
  model-predictive rationality explain network reciprocity over social ties,''
  {\em Scientific Reports}, vol.~9, no.~1, pp.~1--13, 2019.

\bibitem{dellarossa2020reciprocity}
F.~Della~Rossa, F.~Dercole, and A.~Di~Meglio, ``Direct {{Reciprocity}} and
  {{Model-Predictive Strategy Update Explain}} the {{Network Reciprocity
  Observed}} in {{Socioeconomic Networks}},'' {\em Games}, vol.~11, p.~16, Mar.
  2020.

\bibitem{robert1984evolution}
R.~Axelrod, {\em The evolution of cooperation}.
\newblock Basic Books, New York, 1984.

\bibitem{sigmund2010calculus}
K.~Sigmund, {\em The calculus of selfishness}.
\newblock Princeton University Press, 2010.

\bibitem{VonNeumann1944}
J.~von Neumann, O.~Morgenstern, H.~W. Kuhn, and A.~Rubinstein, ``{Theory of
  Games and Economic Behavior},'' {\em Princeton University, Princeton},
  p.~776, 1944.

\bibitem{Myerson1997}
R.~B. Myerson, {\em {Game Theory: Analysis of Conflict}}.
\newblock Harvard University Press, 1~ed., 1997.

\bibitem{szabo2007evolutionary}
G.~Szab{\'o} and G.~Fath, ``Evolutionary games on graphs,'' {\em Physics
  Reports}, vol.~446, no.~4-6, pp.~97--216, 2007.

\bibitem{perc2010coevolutionary}
M.~Perc and A.~Szolnoki, ``Coevolutionary games—a mini review,'' {\em
  BioSystems}, vol.~99, no.~2, pp.~109--125, 2010.

\bibitem{Rapoport1965}
A.~Rapoport and A.~M. Chammah, {\em {Prisoner's dilemma : a study in conflict
  and cooperation}}.
\newblock Ann Arbor: University of Michigan Press, 1965.

\bibitem{szabo2005phase}
G.~Szab{\'o}, J.~Vukov, and A.~Szolnoki, ``Phase diagrams for an evolutionary
  prisoner’s dilemma game on two-dimensional lattices,'' {\em Physical Review
  E}, vol.~72, no.~4, p.~047107, 2005.

\bibitem{eguiluz2005cooperation}
V.~M. Egu{\'\i}luz, M.~G. Zimmermann, C.~J. Cela-Conde, and M.~S. Miguel,
  ``Cooperation and the emergence of role differentiation in the dynamics of
  social networks,'' {\em American journal of sociology}, vol.~110, no.~4,
  pp.~977--1008, 2005.

\bibitem{szolnoki2007cooperation}
A.~Szolnoki and G.~Szab{\'o}, ``Cooperation enhanced by inhomogeneous activity
  of teaching for evolutionary prisoner's dilemma games,'' {\em EPL
  (Europhysics Letters)}, vol.~77, no.~3, p.~30004, 2007.

\bibitem{perc2008social}
M.~Perc and A.~Szolnoki, ``Social diversity and promotion of cooperation in the
  spatial prisoner’s dilemma game,'' {\em Physical Review E}, vol.~77, no.~1,
  p.~011904, 2008.

\bibitem{vilone2014social}
D.~Vilone, J.~J. Ramasco, A.~S{\'a}nchez, and M.~San~Miguel, ``Social imitation
  versus strategic choice, or consensus versus cooperation, in the networked
  prisoner's dilemma,'' {\em Physical Review E}, vol.~90, no.~2, p.~022810,
  2014.

\bibitem{hilbe2015partners}
C.~Hilbe, A.~Traulsen, and K.~Sigmund, ``Partners or rivals? strategies for the
  iterated prisoner's dilemma,'' {\em Games and economic behavior}, vol.~92,
  pp.~41--52, 2015.

\bibitem{javarone2016statistical}
M.~A. Javarone, ``Statistical physics of the spatial prisoner’s dilemma with
  memory-aware agents,'' {\em The European Physical Journal B}, vol.~89, no.~2,
  pp.~1--6, 2016.

\bibitem{hauert2002volunteering}
C.~Hauert, S.~De~Monte, J.~Hofbauer, and K.~Sigmund, ``Volunteering as red
  queen mechanism for cooperation in public goods games,'' {\em Science},
  vol.~296, no.~5570, pp.~1129--1132, 2002.

\bibitem{hauert2002replicator}
C.~Hauert, S.~De~Monte, J.~Hofbauer, and K.~Sigmund, ``Replicator dynamics for
  optional public good games,'' {\em Journal of Theoretical Biology}, vol.~218,
  no.~2, pp.~187--194, 2002.

\bibitem{szabo2002phase}
G.~Szab{\'o} and C.~Hauert, ``Phase transitions and volunteering in spatial
  public goods games,'' {\em Physical Review Letters}, vol.~89, no.~11,
  p.~118101, 2002.

\bibitem{szabo2002evolutionary}
G.~Szab{\'o} and C.~Hauert, ``Evolutionary prisoner’s dilemma games with
  voluntary participation,'' {\em Physical Review E}, vol.~66, no.~6,
  p.~062903, 2002.

\bibitem{vanberg1992rationality}
V.~J. Vanberg and R.~D. Congleton, ``Rationality, morality, and exit,'' {\em
  American Political Science Review}, vol.~86, no.~2, pp.~418--431, 1992.

\bibitem{arend2005modeling}
R.~J. Arend and D.~A. Seale, ``Modeling alliance activity: an iterated
  prisoners' dilemma with exit option,'' {\em Strategic Management Journal},
  vol.~26, no.~11, pp.~1057--1074, 2005.

\bibitem{phelan2005using}
S.~E. Phelan, R.~J. Arend, and D.~A. Seale, ``Using an iterated prisoner's
  dilemma with exit option to study alliance behavior: Results of a tournament
  and simulation,'' {\em Computational \& Mathematical Organization Theory},
  vol.~11, no.~4, pp.~339--356, 2005.

\bibitem{hauert2005vpd}
C.~Hauert and G.~Szab{\'o}, ``Game theory and physics,'' {\em American Journal
  of Physics}, vol.~73, pp.~405--414, May 2005.

\bibitem{chen2010evolutionary}
C.-L. Chen, X.-B. Cao, W.-B. Du, and Z.-H. Rong, ``Evolutionary prisoners
  dilemma game with voluntary participation on regular lattices and scale-free
  networks,'' {\em Physics Procedia}, vol.~3, no.~5, pp.~1845--1852, 2010.

\bibitem{guo2017environment}
H.~Guo, C.~Shen, D.~Dai, M.~Zhang, C.~Chu, and L.~Shi, ``Environment promotes
  the evolution of cooperation in spatial voluntary prisoner's dilemma game,''
  {\em Applied Mathematics and Computation}, vol.~315, pp.~47--53, 2017.

\bibitem{hampton1999environmental}
G.~Hampton, ``Environmental equity and public participation,'' {\em Policy
  Sciences}, vol.~32, no.~2, pp.~163--174, 1999.

\bibitem{buchanan1965economic}
J.~M. Buchanan, ``An economic theory of clubs,'' {\em Economica}, vol.~32,
  no.~125, pp.~1--14, 1965.

\bibitem{imerman1993application}
M.~Imerman {\em et~al.}, ``An application of the theory of club goods to
  capitalization structures and membership investment interests in agricultural
  cooperatives,'' Master's thesis, Iowa State University, 1993.

\bibitem{rogelberg2012wasted}
S.~G. Rogelberg, L.~R. Shanock, and C.~W. Scott, ``Wasted time and money in
  meetings: Increasing return on investment,'' {\em Small Group Research},
  vol.~43, no.~2, pp.~236--245, 2012.

\bibitem{masuda2007participation}
N.~Masuda, ``Participation costs dismiss the advantage of heterogeneous
  networks in evolution of cooperation,'' {\em Proceedings of the Royal Society
  B: Biological Sciences}, vol.~274, no.~1620, pp.~1815--1821, 2007.

\bibitem{szabo1998evolutionary}
G.~Szab{\'o} and C.~T{\H{o}}ke, ``Evolutionary prisoner’s dilemma game on a
  square lattice,'' {\em Physical Review E}, vol.~58, no.~1, p.~69, 1998.

\bibitem{albert2002statistical}
R.~Albert and A.-L. Barab{\'a}si, ``Statistical mechanics of complex
  networks,'' {\em Reviews of Modern Physics}, vol.~74, no.~1, p.~47, 2002.

\bibitem{gomezgardenes2007dynamical}
J.~Gómez-Gardeñes, M.~Campillo, L.~M. Floría, and Y.~Moreno, ``Dynamical
  {Organization} of {Cooperation} in {Complex} {Topologies},'' {\em Physical
  Review Letters}, vol.~98, p.~108103, Mar. 2007.

\bibitem{gomezgardenes2008natural}
J.~Gómez-Gardeñes, J.~Poncela, L.~Mario~Floría, and Y.~Moreno, ``Natural
  selection of cooperation and degree hierarchy in heterogeneous populations,''
  {\em Journal of Theoretical Biology}, vol.~253, pp.~296--301, July 2008.

\bibitem{poncela2007robustness}
J.~Poncela, J.~Gómez-Gardeñes, L.~M. Floría, and Y.~Moreno, ``Robustness of
  cooperation in the evolutionary prisoner's dilemma on complex networks,''
  {\em New Journal of Physics}, vol.~9, pp.~184--184, June 2007.

\bibitem{erdos2011evolution}
P.~Erd{\"o}s and A.~R{\'e}nyi, ``On the evolution of random graphs,'' in {\em
  The Structure and Dynamics of Networks}, pp.~38--82, Princeton University
  Press, 2011.

\end{thebibliography}

\end{document}